\documentclass[12pt]{article}
\begin{document}
\renewcommand{\thefootnote}{\fnsymbol{footnote}}
\begin{center}
{\large\bf STARTING--POINT OF SUPERGRAVITY\\}
\vspace{1cm}
Vyacheslav A. Soroka\footnote{E-mail:  vsoroka@kipt.kharkov.ua}
\vspace{1cm}\\
{\it Institute for Theoretical Physics}\\
{\it National Science Center}\\
{\it "Kharkov Institute of Physics and Technology"}\\
{\it 61108 Kharkov, Ukraine}\\
\end{center}
\vspace{1cm}
\begin{abstract}
In this report I wish to recall how the basic concepts and ingredients of
supergravity were formulated by Dmitrij V. Volkov and the present author
in 1973-74 under the first investigation of the super-Higgs effect .
\end{abstract}
\renewcommand{\thefootnote}{\arabic{footnote}}
\setcounter{footnote}0

\bigskip
\newpage
\section{Introduction}

It is well-known that spontaneously broken gauge symmetries play a
fundamental role for the construction of realistic models for the unified
description of the different interactions (see, for example, \cite{w}).
Gauge symmetries imply the introduction of the massless gauge fields
describing interactions, while spontaneously broken gauge symmetries mean
that some gauge fields become massive, due to the Higgs effect \cite{h},
to describe some kind of forces.  Gauge internal symmetries serve for the
description of the strong and electroweak interactions and gauge
space-time symmetries describe gravity.

The discovery of the extended super-Poincar\'e group \cite{va,av},
nontrivially uniting the space-time and internal symmetries, opened a
principal possibility for the unification of gravity with other
interactions based on the internal symmetries. Since an irreducible
representation of the super-Poincar\'e group, in contrast with the
experience, contains bosons and fermions of the same masses, supersymmetry
has to be broken. Therefore it was reasonable to consider a supersymmetric
unification of gravity and other interactions on the basis of the
spontaneously broken extended super-Poincar\'e gauge group.

The spontaneous breaking of the extended super-Poincar\'e global group
leads to the massless Goldstone fermions with spin 1/2 (goldstinos)
\cite{va,av}.  Real fermionic gauge fields with spin 3/2 (gravitinos),
which appear as superpartners of graviton with spin 2 under gauging of
this group, are necessary ingredients of supergravity theory. The
super--Higgs effect takes place in supergravity when the (extended)
super-Poincar\'e gauge group is spontaneous broken.

If supergravity theory does not contain an interaction with
other (matter) fields, then the super--Higgs effect results in that not
only gravitinos become massive, by absorbing the goldstinos degrees of
freedom, but also the space-time changes its own metric and topological
properties: a nonzero cosmological constant appears. The super--Higgs
effect has been introduced and investigated for the first time in 1973--74
in Refs.  \cite{vs1,vs2}, which in fact can be considered as a
starting--point in the development of supergravity (see also \cite{v}),
since a notion of the super--Higgs effect is meaningless outside the
framework of supergravity.

\section{Spontaneously broken extended super-Poincar\'e\\ gauge group}

N--extended super--Poincar\'e group, whose spontaneous breaking leads
to the appearance of goldstinos, has the following
representation
$$
{G=K(\theta,x)H(l,u)=
\left(\begin{array}{ccc}
1&\theta&ix+{1\over 2}\theta\bar\theta\\
0&1&\bar\theta\\
0&0&1
\end{array}\right)
\left(\begin{array}{ccc}
L&0&0 \\
0&U&0 \\
0&0&(L^+)^{-1}
\end{array}\right)},
\eqno(1)$$
where $\theta^\alpha_k$ is a $2\times n$ matrix with one spinor index
$\alpha$ corresponding to the $SL(2, C)$ group and an index $k$
belongs to a representation of the internal symmetry group given in the
form of the $n\times n$ unitary matrix $U(u)$,
$\bar\theta^{k\dot\alpha}=(\theta^\alpha_k)^+$, $x^{\alpha
\dot\alpha}=x^\mu\sigma_\mu^{\alpha\dot\alpha}$
($\sigma_\mu^{\alpha\dot\alpha}=({\tilde\sigma}_\mu)^{\alpha\dot\alpha}=
(1, -\stackrel{\rightarrow}{\sigma})^{\alpha\dot\alpha}$ are relativistic
Pauli matrices) is a Hermitian $2\times 2$ matrix corresponding to the
translation group and $L(l)$ is a $2\times 2$ matrix of the $SL(2, C)$
group. Quantities $\theta^\alpha_k$ and $\bar\theta^{k\dot\alpha}$, which
will correspond to the goldstino fields,
are Grassmann variables.  Each of the multiplies $K$ and $H$ in (1) form
subgroups and $H$ will be considered as a subgroup of vacuum stability.

The parameters $x$,  $\theta$ and $\bar\theta$ form a homogeneous space
(superspace) with respect to left shifts, i.e., they are transformed among
themselves under a transformation $G_L$
$$
K({\theta}', x')H(l', u') = G_L K(\theta, x)H(l, u).
\eqno(2)$$

The expression $G^{-1}dG$ is invariant under left shifts (2) so that the
Cartan forms, being coefficients of the group generators in the
decomposition
$$
G^{-1}dG = i\omega^\mu{\tilde\sigma}_\mu+\omega_k^\alpha
Q_\alpha^k+{\bar Q}_{k\dot\alpha}{\bar\omega}^{k\dot\alpha}+
\omega^{\mu\nu}(I_{\mu\nu}-I^+_{\mu\nu})+\omega^aI_a,
\eqno(3)$$
are invariant functions under transformations of $G_L$. In (3) the
quantities $Q_\alpha^k$ and ${\bar Q}_{k\dot\alpha}$ are generators
of the spinor translations, $I_{\mu\nu}$ are generators of the $SL(2, C)$
group and $I_a$ $(a=1,...,N)$ are generators of the internal symmetry
group.

For a gauge group $G_L$, which parameters are functions of the point in
superspace with coordinates $z=(x, \theta, \bar\theta)$, the expression (3)
ceases to be an invariant of $G_L$ and acquires an additional
term under the transformations (2) $G'=G_LG$
$$
G'^{-1}dG'=G^{-1}dG+G^{-1}{G_L}^{-1}dG_LG
\eqno(4)$$
For the invariance restoration a gauge differential 1--form $A(d)$ on
superspace has to be introduced
$$
A(d)=\left(
\begin{array}{ccc}
\omega(d)&\psi(d)&ie(d)\\
0        &  V(d) &\psi^+(d)\\
0        &  0    &-\omega^+(d)\\
\end{array}\right)
$$
with the following transformation law
$$
A^\prime(d)=G_LA(d){G_L}^{-1}+G_Ld{G_L}^{-1},
\eqno(5)$$
where $\omega(d)=\omega^{\mu\nu}(d)I_{\mu\nu}$ is a Lorentz connection,
$\psi(d)=\psi^\alpha_k(d)Q_\alpha^k$ and $\psi^+(d)
={\bar Q}_{k\dot\alpha}{\bar\psi}^{k\dot\alpha}$ are Rarita--Schwinger
gauge fields with spin 3/2 (gravitinos), $e(d)=e^\mu(d){\tilde\sigma}_\mu$
is a vierbein 1--form (graviton) and $V(d)=V^a(d)I_a$ are Yang--Mills
gauge fields with spin 1.

Taking into account the transformation rules for the goldstinos (4) and
gauge 1--form (5), it can readily be seen that the differential form
$$
{\cal A}(d)=G^{-1}dG+G^{-1}A(d)G=H^{-1}(dH+\tilde{\cal A}(d)H)
$$
is an invariant of the gauge group $G_L$. Differential 1-form
$$
\tilde{\cal A}(d)=K^{-1}(dK+A(d)K)=\tilde{\cal A}_K(d)+\tilde{\cal A}_H(d)
$$
is not invariant but is transformed as follows
$$
\tilde{\cal A}'_K=H_L\tilde{\cal A}_K{H_L}^{-1},
\eqno(6)$$
$$
\tilde{\cal A}'_H=H_L(d{H_L}^{-1}+\tilde{\cal A}_H{H_L}^{-1}),
\eqno(7)$$
where
$$
\tilde{\cal A}_K=i\tilde e^\mu{\tilde\sigma}_\mu+
\tilde\psi^\alpha_kQ_\alpha^k+
{\bar Q}_{k\dot\alpha}\tilde{\bar\psi}^{k\dot\alpha}
$$
is a part of the gauge 1--form $\tilde{\cal A}(d)$ which belong to the
class $K$ and
$$
\tilde{\cal A}_H=A_H=\omega(d)-\omega^+(d)+ V(d)
$$
is a part of the form $\tilde{\cal A}(d)$ which belong to the subgroup $H$.

Entering into $\tilde{\cal A}_K(d)$ 1--forms are
$$
\tilde e(d)=e(d)+Dx+{i\over2}\left[\theta(2\bar\psi(d)+D\bar\theta)-
(2\psi(d)+D\theta)\bar\theta\right],
\eqno(8)$$
$$
\tilde\psi(d)=\psi(d)+D\theta, \qquad
\tilde{\bar\psi}(d)=\bar{\psi}(d)+D\bar\theta,
\eqno(9)$$
where covariant derivatives have the following forms
$Dx=dx+\omega(d)x+x\omega^+(d)$,
$D\theta=d\theta+\omega(d)\theta-\theta V(d)$ and
$D\bar\theta=d\bar\theta+\bar\theta\omega^+(d)+V(d)\bar\theta$.

To construct kinetic terms for the gravitinos, graviton and Yang--Mills
fields it will be necessary to use the following covariant 2--forms
$$
D(d)\tilde\psi(\delta)=d\tilde\psi(\delta)+
\omega(d)\tilde\psi(\delta)-\tilde\psi(\delta)V(d),
\eqno(10)$$
$$
R^{\mu\nu}(d,\delta)=d\omega^{\mu\nu}(\delta)-\delta\omega^{\mu\nu}(d)+
2\left[{\omega^\mu}_\lambda(d)\omega^{\lambda\nu}(\delta)-
{\omega^\nu}_\lambda(d)\omega^{\lambda\mu}(\delta)\right],
\eqno(11)$$
$$
F^a(d,\delta)=dV^a(\delta)-\delta V^a(d)+{c_{bc}}^aV^b(d)V^c(\delta),
\eqno(12)$$
where ${c_{bc}}^a$ are structure constants for the group of internal
symmetry.
It can easily be verified that 2--forms (10)--(12) are transformed under
the gauge group $G_L=K_LH_L$ homogeneously by means of the subgroup
transformation $H_L$ as well as the 1--form $\tilde{\cal A}_K(d)$ is
transformed in accordance with (6).

\section{Invariant action and the super--Higgs effect}

The invariant action integral is derived from the expressions (8)--(12) as
a sum with arbitrary constants of exterior products of forth order (or in
the general case in the form of a homogeneous function of first order of
such products) that are invariant under the Lorentz group and the internal
symmetry group.

The simplest admissible invariant combinations, corresponding to the
requirement that the degree of the field derivatives be minimal, are
$$
I_1=iR_{\dot\alpha}^{\dot\beta}(d_1,d_2)\wedge
{\tilde e}_{\dot\beta\gamma}(d_3)\wedge{\tilde e}^{\gamma\dot\alpha}(d_4)+
h.c. ,
\eqno(13)$$
$$
I_2=D(d_1)\wedge{\tilde\psi}_k^\alpha(d_2)\wedge{\tilde
e}_{\alpha\dot\beta}(d_3)\wedge{\tilde{\bar\psi}}^{k\dot\beta}(d_4)+h.c. ,
\eqno(14)$$
$$
I_3={\tilde\psi}_k^\alpha(d_1)\wedge{\tilde e}_{\alpha\dot\beta}(d_2)\wedge
{\tilde e}^{\dot\beta\gamma}(d_3)\wedge{\tilde\psi}_\gamma^k(d_4)+h.c. ,
\eqno(15)$$
$$
I_4=i{\tilde e}_\alpha^{\dot\beta}(d_1)\wedge
{\tilde e}_{\dot\beta}^\gamma(d_2)
\wedge{\tilde e}_\gamma^{\dot\delta}(d_3)\wedge
{\tilde e}_{\dot\delta}^\alpha(d_4) ,
\eqno(16)$$
$$
I_5=\left[F^a(d_1,d_2)\wedge{\tilde e}_\alpha^{\dot\beta}(d_3)\wedge
{\tilde e}_\gamma^{\dot\delta}(d_4)\right]
\left[F_a(d_5,d_6)\wedge{\tilde e}_{\dot\beta}^\alpha(d_7)\wedge
{\tilde e}_{\dot\delta}^\gamma(d_8)\right]\times(I_4)^{-1} .
\eqno(17)$$
The quantities in (13)--(17) are given in the spinor representation a
transition to which from the tensor representation is
$$
e^{\alpha\dot\beta}=(\tilde\sigma_\mu)^{\alpha\dot\beta} e^\mu , \qquad
{R^\alpha}_\beta={({\tilde\sigma}^\mu\sigma^\nu)^\alpha}_\beta R_{\mu\nu} ,
\qquad
{R_{\dot\beta}}^{\dot\alpha}={(\sigma^\mu{\tilde\sigma}^\nu)_{\dot\beta}}^
{\dot\alpha}R_{\mu\nu} ,
$$
where ${\sigma^\mu}_{\dot\alpha\beta}=
(1,-\stackrel{\rightarrow}{\sigma})_{\dot\alpha\beta}$.

The invariant expression (13) contains as a term the Einstein--Cartan
action for the gravitational field while the invariant (17) has a term for
the Yang--Mills fields. The expression (16) contains the kinetic term for
the goldstinos. The invariants (14) and
(15) include respectively the kinetic and mass terms for
Rarita--Schwinger real (Majorana) fields with spin 3/2 (gravitinos).

When all the field variables are parametrized as functions of $x$, all the
gauge differential forms $A(d)$ can be reduced by a redefinition of the
fields to the form
$$
A(d)=A_\mu(x)dx^\mu ,
$$
where $A_\mu(x)$ are the redefined gauge fields.

In order to study the Higgs effect contained in the expressions
(13)--(17) it is necessary to examine the structure of the Cartan forms
in the definitions (8)--(12).

By a transition with the help of the left shift $G_L=K^{-1}(x,\theta)$ to
the gauge in which the evident dependence on the goldstinos
$\theta$ and space-time coordinates $x$ disappears, the transformed gauge
fields become
$$
\psi'(d)=\tilde\psi(d),\qquad e'_\mu(d)={\tilde e}_\mu(d).
\eqno(18)$$
In this gauge the following conditions $\theta'=0$ and $x'=0$ are realized.
This conditions are invariant under gauge transformations from the
subgroup $G_L=H(l,u)$ and therefore do not fix the gauge completely.
Under the rest gauge transformations from the subgroup $G_L=H(l,u)$ of
vacuum stability the forms $\tilde{\cal A}_K(d)$, which contain the forms
$\psi(d)$ and $e_\mu(d)$, absorbing the goldstinos degrees of freedom and
transforming homogeneously accordingly to (6), lost its gauging nature,
while the forms $\tilde{\cal A}_H$, transforming in
accordance with (7) inhomogeneously, remain truly gauge forms.

As a result of the redefinition (18), the invariants (13)--(17)
correspond to the interaction of the gravitational field with cosmological
term (16), massive gravitinos, and Yang--Mills
fields. That corresponds to the super--Higgs effect, as a result of which
the goldstinos disappear because of the redefinition (18) of the
quantities associated with the gravitinos and metric and topological
properties of space-time.  Moreover, in contrast with the Higgs effect for
the group of internal symmetry, the kinetic term for the goldstinos
transits not into the mass term of gravitinos but into the cosmological
term, while the mass term of gravitinos appears from the term which in the
absence of gauge fields is a total derivative.

Thus, a mechanism of the super--Higgs effect essentially differs than
the Higgs effect for the Goldstone particles with spin zero which takes
place in the case of a spontaneously broken group of internal symmetry.

\section{Conclusion}

Thus, we see that the main notions and ingredients of supergravity were
formulated in our papers \cite{vs1,vs2}.

1. It has been shown that the extended super-Poincar\'e gauge group
leads to the supersymmetric generalization of gravity which has to include
the real gauge fields with spin 3/2 as
graviton superpartners. This fact was very unusual for those times,
because till then the gauge fields possessed only integer spins.  The
graviton superpartners with spin 3/2 were called later on as gravitino
fields.

2. We have written down our action as a sum of the five invariant terms
$I_1,...,I_5$ (13)--(17) with arbitrary constants before them. The
invariants $I_1$ and $I_2$ contain respectively: the Einstein--Cartan
action for gravity and a kinetic term for the gravitino fields which enter
into the action for pure supergravity. The relation between the
constants before the terms $I_1$ and $I_2$ has been established as a
result of the powerful progress achieved by Ferrara, Freedman, van
Nieuwenhuizen \cite{ffn} and Deser, Zumino \cite{dz} under construction of
the on--shell formulation for $N=1$ supergravity.

The invariant terms $I_3$ and $I_4$, containing correspondingly a
mass term of the gravitinos and a cosmological term, appear as a result of
the super--Higgs effect.

3. The term $I_5$ for the Yang--Mills fields with spin 1 demonstrates a
most important fact that the extended super-Poincar\'e gauge group gives a
principle possibility for the nontrivial unification of gravity with the
interactions based on the internal symmetries.

As the complete list of papers on the theme is very huge
and essentially exceeds this report, I refer only to our works and to those
very closely related to them. A very extensive list of references
concerning the development of supergravity can be found in Ref. \cite{n}.

\end{document}